# Probability and the Classical/Quantum Divide

Subhash Kak[1]

## ABSTRACT

This paper considers the problem of distinguishing between classical and quantum domains in macroscopic phenomena using tests based on probability and it presents a condition on the ratios of the outcomes being the same ($P_s$) to being different ($P_n$). Given three events, $P_s/P_n$ for the classical case, where there are no 3-way coincidences, is one-half whereas for the quantum state it is one-third. For non-maximally entangled objects $|\varphi\rangle_{AA} = \frac{1}{\sqrt{1+r^2}}(r|00\rangle + |11\rangle)$ we find that so long as $r < 5.83$, we can separate them from classical objects using a probability test. For maximally entangled particles ($r = 1$), we propose that the value of 5/12 be used for $P_s/P_n$ to separate classical and quantum states when no other information is available and measurements are noisy.

## INTRODUCTION

That a signal be considered quantum is generally determined by characteristics such as superposition, entanglement, and complementarity. Information associated with either classical or quantum variables is measured in relation to the experimental arrangement and it is a function of the probabilities of the outcomes associated with the experiment. When calculating information, one should first determine whether the variable being measured is classical or quantum but this may be difficult if the measurement constraints preclude determination of, say, entanglement.

If the variable is quantum, one must determine if it is pure or mixed and this would require prior testing and assumption of stationarity of the process with respect to time. One must also define the framework within which the question of information is being asked since, under certain conditions, an unknown pure state – associated with zero von Neumann entropy -- can convey information [1].

There are many physical processes where the nature of the variables is well established both by theory and experiment. But what about new situations, outside of physics in macroscopic systems, where there isn't consensus that the processes have a quantum mechanical basis? Quantum biology is one such area [2]-[5]. Reasons have also been advanced for considering quantum models of the brain [6]-[14] and if such functioning is true there ought to be evidence in favor of coincidences across space and time [15].

In many situations, such as determination of entanglement, the differentiation between classical and quantum effects is estimated by checking if the Bell inequality is violated [16]-[18]. In his original paper [16], Bell showed that under conditions of independence classical random variables A, B, C will satisfy

$$|P_{same}(A,B) - P_{same}(A,C)| \leq P_{same}(B,C) + 1 \tag{1}$$

---

[1] Department of Computer Science, Oklahoma State University, Stillwater, OK 74074.

where $P_{same}(X,Y)$ is the probability that the pair of random variables X, Y have some identical property. The Bell inequality is a consistency constraint on functions of two random variables and it can be described in many other ways (and we will use a slightly different form in our discussion). Bell showed that similar measurement of entangled quantum variables can lead to a violation of the inequality and, therefore, such violation can serve as a divide between classical and quantum variables. But Bell inequality is also violated in many classical situations where long-range correlations persist [19]-[22].

Statistics may also be used to distinguish between classical and quantum systems [23]. But this applicability will be limited to situations where the quantum process is in a state of thermal equilibrium. When a closed quantum system with a large number of eigenstates is opened to the environment, many of these states couple to the environmental states and undergo decoherence. This creates environment-induced superselection that leads to the survival of certain states [24] that remain correlated with the universe, obeying classical statistics, even though they are quantum mechanical.

The classical/quantum divide is a central notion of the Copenhagen Interpretation (CI) of quantum mechanics [25],[26]. The Many Worlds Interpretation (MWI) takes the wavefunction to be the primary reality and assigns a wavefunction to the universe itself. In MWI there is no collapse of the wavefunction and the observation is a consequence of decoherence brought about by the environment. If CI is an inside-out view of the universe where the reality is constructed out of the perceptions of the experimenter, MWI is an outside-in view in which the mathematical function of the universe is the primary reality [27]. In our view CI is better than MWI in addressing the question of free-will which it does through the quantum Zeno effect if consciousness is seen to have a universal basis [28].

In application of quantum theory to beliefs (which sidesteps the question whether the brain must be described by a quantum model), questions A, B, A|B, and B|A are presented in sequence to each member of a group of people and the responses averaged [29],[30]. If it is shown that P(A|B) P(B) ≠ P(B|A) P(A), that would be evidence in support of a non-classical basis to such probability. In entangled states such as $|\varphi\rangle = 0.9|00\rangle - 0.3|01\rangle - 0.1|10\rangle - 0.3|11\rangle$, event order leads to different results for P(01) = 0.09 and P(10) = 0.01. Such an entangled state may be created by passing the state $|00\rangle$ through some appropriate unitary operator although it is not clear if it is easy to implement arbitrary operators [31]. Six reasons have been advanced for a quantum approach to mental events [29]: mental states are indefinite, judgments create rather than merely record, they disturb each other, and they do not always obey classical logic, they do not obey the principle of unicity, and cognitive phenomena may not be decomposable. To set up a proper quantum analogy for a questioning-answering system, we must assume that there is an ideal state function, in a suitable Hilbert space, that has been created by society and individuals are filters who change their state with a certain probability to that corresponding to the filter orientation.

We take another look at what is possible to infer from experimental observations related to the nature of the system or the signals. Given pairwise data for variables, Boole showed what constraints had to be satisfied for the pairwise probabilities of a set of three evens to be consistent [32]. In this sense, Boole's work prefigures Bell's inequalities [16] although the focus of Bell was to find a setting that would clearly point to the differences between classical and quantum situations. We use the Bell-type inequality [33] on the probability of same outcomes for pairwise consideration of three pairs of bases.

An entangled quantum state such as $|\varphi\rangle = \frac{1}{\sqrt{2}}(|00\rangle + |11\rangle)$ has the same form along all measurement bases. Thus for any random bases $|b_0\rangle, |b_1\rangle$ $|\varphi\rangle = \frac{1}{\sqrt{2}}(|b_0 b_0\rangle + |b_1 b_1\rangle)$ and each of the qubits is in a mixed state. It is this feature that makes the measurements along different bases on the quantum state different from the measurements of classical states.

The conditions for the probability constraints are well-defined both for classical objects and quantum states and we stress that the use of the ratios of "same" to "nonsame" events will be more reliable in the presence of noise than to consider just the absolute value of "same" events. For non-maximally entangled objects $|\varphi\rangle_{AA} = \frac{1}{\sqrt{1+r^2}}(r|00\rangle + |11\rangle)$ we find that so long as $r < 5.83$, we can separate them from classical objects based on computation of "same" and "nonsame" events across appropriately chosen bases.

## PROBABILITY CONSTRAINTS

It is quite clear that a constraint on n variables will be projected as several constraints on subsets of these variables. Therefore, if we are only given the constraints on the subsets of the variables, it cannot be said if they were derived from the same function on all the variables. Consider (with Boole) three events A, B, and C. Let $P(AB) = r$; $P(BC) = s$; and $P(AC) = t$. If a Venn diagram is drawn and we write

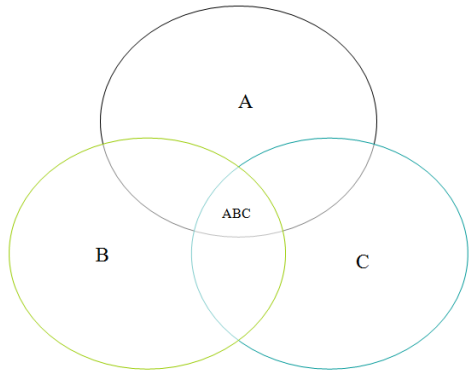

Figure 1. Venn diagram for three events A, B, C

$$P(ABC) = \lambda; P(AB\overline{C}) = \mu; P(A\overline{B}C) = v; P(\overline{A}BC) = \eta$$

Then
$$\lambda + \mu = r; \lambda + \eta = s; \lambda + v = t \qquad (2)$$

A straightforward computation shows that the following constraints need to be satisfied for the data to be consistent:

$$r \geq s + t - 1$$
$$s \geq t + r - 1 \qquad (3)$$
$$t \geq r + s - 1$$

These are of the form

$$P(AB) - P(AC) \geq P(BC) - 1$$

which may be rewritten as:

$$|P(AB) - P(AC)| \leq 1 - P(BC) \qquad (4)$$

This is a form similar to that of the original Bell inequality (1).

For example, the values of 2/5, 2/3, and 4/5 for *r, s,* and *t* will not constitute a consistent set as the inequality (4) is violated. The source of the inconsistency may be that the data is corrupt or the assumption that the objects being considered in the sample space are classical (or fixed in properties) is incorrect.

Now we consider a more general situation where there are three random variables A, B, and C each one of which takes two values, say 0 and 1. Let the probabilities of the eight possible outcomes be defined as below:

Table 1. Probability of outcomes for events

| A | B | C | Probability |
|---|---|---|---|
| 0 | 0 | 0 | a |
| 0 | 0 | 1 | b |
| 0 | 1 | 0 | c |
| 0 | 1 | 1 | d |
| 1 | 0 | 0 | e |
| 1 | 0 | 1 | f |
| 1 | 1 | 0 | g |
| 1 | 1 | 1 | h |

In real data, the set may not be complete due to noise and other experimental errors. Clearly,

$$P_{AB}(0,0) = a + b; P_{AB}(0,1) = c + d; P_{AB}(1,0) = e + f; P_{AB}(1,1) = g + h$$

$$P_{BC}(0,0) = a + e; P_{BC}(0,1) = b + f; P_{BC}(1,0) = c + g; P_{BC}(1,1) = d + h$$

$$P_{AC}(0,0) = a + c; P_{AC}(0,1) = b + d; P_{AC}(1,0) = e + g; P_{AC}(1,1) = f + h$$

Since these are mutually dependent conditions, they should satisfy conditions such as:

$$P_{AB}(0,0) + P_{AB}(0,1) = P_{AC}(0,0) + P_{AC}(0,1) \tag{5}$$

$$P_{BC}(0,0) + P_{BC}(0,1) = P_{AB}(0,0) + P_{AB}(1,0) \tag{6}$$

$$P_{BC}(0,0) + P_{BC}(1,0) = P_{AC}(0,0) + P_{AC}(1,0) \tag{7}$$

that express $p_i(0)$ in two different ways. Likewise, there would be a further set of conditions that express $p_i(1)$ in two different ways.

Let us now define $P_{AB}(0,0) + P_{AB}(1,1) = P_{same}(A,B)$, $P_{AC}(0,0) + P_{AC}(1,1) = P_{same}(A,C)$, and $P_{BC}(0,0) + P_{BC}(1,1) = P_{same}(B,C)$. It follows that the following result is true:

$$P_{same}(A,B) + P_{same}(A,C) + P_{same}(B,C) = 1 + 2a + 2h \tag{8}$$

**Theorem 1**. Consider that a sample space is completely described by the binary outputs associated with events A, B, and C. Then $P_{same}(A,B) + P_{same}(A,C) + P_{same}(B,C) = 1 + 2P_{same}(A,B,C)$ where $P_{same}(A,B,C) = P_{ABC}(0,0,0) + P_{ABC}(1,1,1)$.

Theorem 1 implies the general (Bell-type) inequality:

$$P_{same}(A,B) + P_{same}(A,C) + P_{same}(B,C) \geq 1 \tag{9}$$

It is obvious that the satisfaction of the inequality (9) is necessary but not sufficient for the data to be consistent. Consider the probability values given below for three random variables A, B, C:

$$P_{AB}(0,0) = 0.3; P_{AB}(0,1) = 0.2; P_{AB}(1,0) = 0.3; P_{AB}(1,1) = 0.2$$

$$P_{BC}(0,0) = 0.15; P_{BC}(0,1) = 0.35; P_{BC}(1,0) = 0.25; P_{BC}(1,1) = 0.25 \tag{10}$$

$$P_{AC}(0,0) = 0.1; P_{AC}(0,1) = 0.4; P_{AC}(1,0) = 0.2; P_{AC}(1,1) = 0.3$$

These values violate the inequalities (5), (6), and (7). The data does however satisfy the inequality (9) since the left hand side for this case is 0.5+0.4+0.4 > 1.

Now consider a specific case of three independent variables associated with an experiment where each variable takes the values 0 or 1. Let the probabilities of A, B, and C being 1 and 0 be given by the table below:

|      | A   | B   | C   |
|------|-----|-----|-----|
| p(1) | r   | s   | t   |
| p(0) | 1-r | 1-s | 1-t |

Consider the product:

$$rs + (1-r)(1-s) + st + (1-s)(1-t) + rt + (1-r)(1-t) \tag{11}$$

To find its minimum we differentiate this expression with respect to r and put that equal to zero. We find that $s = -t + 1$. Substituting back in (11) we find the minimum to be $1 + 2t - 2t^2$ whose least value is 1. Its maximum value is obtained when r, s, t are each equal to 1 which reduces equation (11) to 3. The left hand side of (11) is nothing but $P_{same}(A,B) + P_{same}(A,C) + P_{same}(B,C)$ and, therefore, we can write:

$$P_{same}(A,B) + P_{same}(A,C) + P_{same}(B,C) \geq 1 \tag{12}$$

Let $P_{sc} = P_{same}(A,B) + P_{same}(A,C) + P_{same}(B,C)$, and $P_{nc} = P_{notsame}(A,B) + P_{notsame}(A,C) + P_{notsame}(B,C)$ where $P_{notsame}(A,B)$ means that the outputs A and B are different, etc, and the "c" in the subscript for $P_{sc}$ and $P_{nc}$ refers to "classical" for what is being considered is classical probability. A straightforward computation shows that

$$P_{nc} = P_{notsame}(A,B) + P_{notsame}(A,C) + P_{notsame}(B,C) = 2b + 2c + 2d + 2e + 2f + 2g = 2 - 2P_s(A,B,C) \tag{13}$$

By using Theorem 1 on (13), it follows that

$$P_{sc} + P_{nc} = 3 \tag{14}$$

We can summarize the result on the ratio of the probabilities of getting "same" output to "notsame" output by the next result:

**Theorem 2.** $\dfrac{P_{sc}}{P_{nc}} = \dfrac{1 + 2P_s(A,B,C)}{2[1 - P_s(A,B,C)]}$ (15)

In the special case where $P_{same}(A,B,C) = 0$,

$$\frac{P_{sc}}{P_{nc}} = \frac{1}{2} \tag{16}$$

## CORRELATION ACROSS TWO DIFFERENT BASES

We wish to determine the correlation across two bases for the general quantum case of entangled particles

$$|\varphi\rangle_{AA} = \frac{1}{\sqrt{1+r^2}}(r|00\rangle + |11\rangle) \tag{17}$$

originally described with respect to the A set of bases:

$$|a_0\rangle = |0\rangle, |a_1\rangle = |1\rangle. \tag{18}$$

Let the second set of bases be

$$|b_0\rangle = \frac{1}{\sqrt{k}}|0\rangle - \frac{\sqrt{k-1}}{\sqrt{k}}|1\rangle, |b_1\rangle = \frac{\sqrt{k-1}}{\sqrt{k}}|0\rangle + \frac{1}{\sqrt{k}}|1\rangle \tag{19}$$

Substituting, we see that

$$|\varphi\rangle_{BB} = \frac{1}{\sqrt{1+r^2}}\{\frac{(r+(k-1))}{k}|b_0 b_0\rangle - \frac{\sqrt{k-1}}{k}(r-1)|b_0 b_1\rangle - \frac{\sqrt{k-1}}{k}(r-1)|b_1 b_0\rangle + \frac{r(k-1)+1}{k}|b_1 b_1\rangle\} \tag{20}$$

For the maximally entangled cased of $r=1$, $|\varphi\rangle_{BB} = \frac{1}{\sqrt{2}}(|b_0 b_0\rangle + |b_1 b_1\rangle)$.

To consider the correlation across the two bases, we consider $|\varphi\rangle_{AB}$ representing the projection of one set of bases on another. We can express the state $|\varphi\rangle_{AB}$ as:

$$|\varphi\rangle_{AB} = \frac{1}{\sqrt{k}\sqrt{1+r^2}}(r|00\rangle - r\sqrt{k-1}|01\rangle + \sqrt{k-1}|10\rangle + |11\rangle) \tag{21}$$

The probabilities $P_{same}(A,B)$ and $P_{notsame}(A,B)$ will simply be the modulus square values of the amplitudes of $|00\rangle$ and $|11\rangle$ on the one hand, and $|01\rangle$ and $|10\rangle$ on the other. Making the required computation, we obtain:

$$P_{same}(A,B) = \frac{1}{k} \text{ and } P_{notsame}(A,B) = \frac{k-1}{k} \tag{22}$$

**Theorem 3.** For the non-maximally entangled particles $|\varphi\rangle_{AA} = \frac{1}{\sqrt{1+r^2}}(r|00\rangle + |11\rangle)$ and the two frames defined by (18) and (19), the probabilities of same output to dissimilar output is given by

$$\frac{P_{same\_quantum}(A,B)}{P_{notsame\_quantum}(A,B)} = \frac{1}{k-1} \qquad \blacksquare \tag{23}$$

*It is significant that the result is independent of the value of r.* This means that we will get the same result even if the entanglement is zero and we are dealing with a pure state.

The expression (23) gives the same probabilities for similar and dissimilar outputs for $k = 2$. This is the case which corresponds to measurement bases $|b_i\rangle$ that are orthogonal to $|a_i\rangle$. As $k$ is increased, we rotate $|b_0\rangle$ away from $|a_0\rangle$. The value $k=2$ corresponds to $-45^0$, $k=3$ to $-54^0$, and $k=4$ to $-60^0$, and so on.

A straightforward calculation shows that for the state $|\varphi\rangle_{AA} = a|00\rangle + b|01\rangle + c|10\rangle + d|11\rangle$ a correlation across the bases BB of (19) leads to the following $P_{same}(A,B)$:

$$\frac{a^2 + b^2(k-1) + c^2(k-1) + d^2}{k} \tag{24}$$

## LOCALITY VERSUS NONLOCALITY

The context for the Bell inequalities is the EPR experiment in which two observers make measurements on a pair of entangled objects. Observers A (Alice) and B (Bob) make their observations using filters whose orientations are chosen independently immediately before their observation and, therefore, these orientations cannot be considered to be related. For an entangled pair of objects, if the measurement orientations chosen by Alice and Bob happen to be the same, the results of the filters will be identical. Since the probability of each of the two choices being a 0 or 1 is equal, this implies that somehow the two objects remained coupled even though they could be so far apart that signal from one could not reach the other in the time that the decision to use a specific measurement orientation was made.

In an actual experiment, the measurement orientations will be randomly distributed. Nevertheless, the Bell inequalities describe the situation if the attributes of the two objects are defined in advance of the measurement and there are no existing correlations between the separated objects. A purely quantum description, on the other hand, violates the Bell inequalities.

An essential ingredient of the argument related to the experiment is that the actions of the source of the entangled objects cannot possibly depend on the later decisions related to measurement orientations made by both A and B. One may say that free will was behind these decisions. Parenthetically, it should be noted that the maximally entangled pair of objects are not in superposition states but are rather mixtures. The measurement by Alice does not cause a collapse; it is a choice between two outcomes associated with equal probability.

The violation of Bell inequalities has been generally viewed as the best proof of nonlocality. But all experiments that have been conducted until now have loopholes and therefore they cannot be considered as definitive tests. The two principal loopholes are those of *locality* and *detection*. The locality loophole addresses the possibility that a local realistic theory might rely on some type of signal sent from one entangled particle to its partner. The detection loophole addresses the fact that when measured with low-quantum-efficiency detectors, as is currently the case, experimental results can be explained by a local realistic theory [33]-[36].

The inequality (9) is violated by measurements on certain quantum states. Thus if the state function is the non-maximally entangled state $|\varphi\rangle_{AA} = \frac{1}{\sqrt{1+r^2}}(r|00\rangle + |11\rangle)$, we can take the three measurement bases to be the $0^0$, $90^0$; $-60^0$, $30^0$ ($k=4$) and $-30^0$, $60^0$ ($k=4$ and flipping the frames) to maximize the total probability by picking a uniform distribution over the phase space (as in [29]). These frames are written as:

$$|a_0\rangle = |0\rangle, |a_1\rangle = |1\rangle \tag{25}$$

$$|b_0\rangle = \frac{1}{2}|0\rangle - \frac{\sqrt{3}}{2}|1\rangle, |b_1\rangle = \frac{\sqrt{3}}{2}|0\rangle + \frac{1}{2}|1\rangle \tag{26}$$

$$|c_0\rangle = \frac{1}{2}|0\rangle + \frac{\sqrt{3}}{2}|1\rangle, |c_1\rangle = \frac{\sqrt{3}}{2}|0\rangle - \frac{1}{2}|1\rangle \tag{27}$$

We have already seen in equation (22) that $P_{same}(A,B) = \text{Prob}(|00\rangle + |11\rangle) = 1/k$ which is ¼. Similarly, $P_{same}(A,C) = ¼$.

The function $|\varphi\rangle_{BC}$ may be derived from (17) by substituting the inverse transformations for (26) and (27):

$$|\varphi\rangle_{BC} = \frac{1}{\sqrt{1+r^2}}\{\frac{(r-3)}{4}|b_0 c_0\rangle + \frac{\sqrt{3}}{4}(r+1)|b_0 c_1\rangle - \frac{\sqrt{3}}{4}(r+1)|b_1 c_0\rangle + \frac{3r-1}{4}|b_1 c_1\rangle\} \tag{28}$$

Therefore,

$$P_{same}(B,C) = \frac{1}{(1+r^2)}\{\frac{(r-3)^2}{16} + \frac{(3r-1)^2}{16}\} = \frac{5r^2 - 6r + 5}{8(1+r^2)} \tag{29}$$

Therefore, we find that in the quantum example considered above,

$$P_{sq} = P_{same}(A, B) + P_{same}(A, C) + P_{same}(B, C) = \frac{1}{2} + \frac{5r^2 - 6r + 5}{8(1+r^2)} \tag{30}$$

Equation (30) violates the Bell-type inequality (9), that is it is less than 1, unless $r \geq 5.83$. It is equal to 1 when $r^2 - 6r + 1 = 0$, which has solutions $3 \pm \sqrt{8}$ or 5.83 and 0.17 which are the same solution since they are inverses of each other. Furthermore,

$$P_{nq} = P_{notsame}(A,B) + P_{notsame}(A,C) + P_{notsame}(B,C) = \frac{15r^2 + 6r + 15}{8(1+r^2)} \tag{31}$$

When $r = 1$, $P_{same}(B,C) = ¼$ and so in each of these cases, the "notsame" probability is ¾.

Summarizing, we can write:

**Theorem 4**. For the state $|\varphi\rangle_{AA} = \frac{1}{\sqrt{1+r^2}}(r|00\rangle + |11\rangle)$, for $P_{sq}$ and $P_{nq}$ defined as in (27) and (28), $P_{sq} < 1$ for $r < 5.83$. We have the further result:

$$\frac{P_{sq}}{P_{nq}} = \frac{3r^2 - 2r + 3}{5r^2 + 2r + 5} \qquad (32)$$

The probability of obtaining the "same" results in the quantum case (32) is lower than for the classical case (15) for $r < 5.83$.

## CONCLUSIONS

This paper considers the problem of distinguishing between classical and quantum domains in macroscopic phenomena using tests based on probability and it presents a condition on the ratios of the outcomes being the same ($P_s$) to being different ($P_n$). Given three events, $P_s/P_n$ for the classical case where there are no three-way coincidences is one-half whereas for the maximally-entangled quantum state it is one-third. In the presence of noise, the middle point of 5/12 could serve as the dividing line between these two states. For the non-maximally entangled state $\frac{1}{\sqrt{1+r^2}}(r|00\rangle + |11\rangle)$, the three-way coincidences can distinguish between quantum and classical cases so long as $r < 5.83$.

It is indeed true that the conditions outlined here will also be satisfied by many regimes of classical waves [19],[37]. Clearly the applicability of these probability constraints is relevant when we are considering the particle view of the phenomenon.

*Acknowledgements.* This research was supported in part by the National Science Foundation grant CNS-1117068. I am also thankful to the Chopra Foundation for hospitality at the Sages and Scientists Symposium 2013 where this work was completed.